\documentclass[journal=jpclcd,manuscript=article,tocentry=none]{achemso}
\makeatletter
\renewcommand{\acs@tocentry@print}[1]{}
\renewcommand{\acs@tocentry@print@aux}{}
\@ifpackageloaded{mciteplus}{\mciteErrorOnUnknownfalse}{}
\makeatother
\usepackage[T1]{fontenc} 
\usepackage{graphicx}
\usepackage[normalem]{ulem}
\usepackage{array}
\usepackage{booktabs}
\usepackage{subcaption}
\usepackage{amsmath}
\usepackage{dirtytalk}
\usepackage{color}
\usepackage{amsfonts}
\usepackage{comment}
\usepackage{xr}
\usepackage{xcolor} 
\usepackage{bbm}
\usepackage{caption}

\title{Model structures and electron transfer properties of conductive nickel-organic nanoribbons in cable bacteria}
\author{Oliver Russell}
\affiliation[UCL]{Department of Physics and Astronomy and Thomas Young Centre, University College London, London, WC1E 6BT, United Kingdom}
\author{Martijn A. Zwijnenburg}
\affiliation[UCL]{Department of Chemistry and Thomas Young Centre,
University College London, London, WC1E 6BT, United Kingdom}
\author{Filip J.R. Meysman}
\affiliation[Antwerp]{Department of Biology, University of Antwerp, Antwerp, Belgium}
\author{Jochen Blumberger}
\email{j.blumberger@ucl.ac.uk}
\affiliation[UCL]{Department of Physics and Astronomy and Thomas Young Centre, University College London, London, WC1E 6BT, United Kingdom}

\begin{document}

\maketitle

\newpage

\begin{abstract}
Cable bacteria are multicellular bacteria capable of centimeter-scale conduction through a regular fiber network embedded in their cell envelope. The conductivity of these fibers is extremely high for biological materials, and rivals that of the best synthetic conductive polymers, but the underlying electron transport mechanism remains elusive. Recent microscopic and spectroscopic evidence indicates that each fiber embeds a bundle of intertwined nanoribbons as the conductive conduit. Each nanoribbon consists of a one-dimensional nickel-organic framework, built from stacked nickel bis(1,2-dithiolene) oligomers (NiBiD units) as molecular building blocks. Here we performed DFT calculations of nanoribbon model structures, in order to characterize their electronic properties, examine potential stacking configurations and verify whether these structures can support efficient conductance. Our simulations indicate that nanoribbons are comprised of tightly stacked  AA or AB-type packings of NiBiD units. In the most energetically stable structure (AB-type) some Ni centers are predicted to be 5-fold coordinated due to formation of an inter-layer Ni-S coordination bond. In several energetically low-lying structures, the electronic coupling between neighboring molecules exceeds the critical threshold for charge delocalization permitting efficient charge transport beyond small polaron hopping. Our results hence reveal that nanoribbons based on NiBiD units exhibit favorable charge transfer properties that may explain the unusually high conductivities measured in the fibers of cable bacteria.

\end{abstract}

\newpage
Electron transfer (ET) is at the heart of virtually all energy conversion processes in living  organisms, with photosynthesis and respiration providing the most prominent examples\cite{Moser92,Gray03}. At the cellular level, ET is realized over nanometer length scales by assemblies of proteins that host redox-active cofactors, typically hemes\cite{Hartshorne09,Wang19} or Fe-S clusters\cite{Zhu16,Gao25}. These cofactors are meticulously arranged into chains within the protein matrix, separated by distances of < 2 nm, where they act as stepping stones for the electrons to traverse the insulating protein. This electron "hopping" process is generally well described by the Marcus theory of non-adiabatic electron transfer \cite{Blumberger15}. The respiratory complexes of the mitochondrial membrane\cite{Zhu16,Brzezinski21} as well as the multiheme cytochrome complexes of rock-respiring bacteria\cite{Hartshorne09,Subramanian18} are key examples of this natural design principle. 

Over the past two decades, it has however become clear that the spatial reach of biological electron transport may far exceed the nanometer length scale \cite{Subramanian18,Wang19,Malvankar12}. The most striking example of such extremely long-range biological charge transport are cable bacteria. These multicellular bacteria are found in freshwater and marine sediments and capable of transferring electrons over 
centimeter scales\cite{Nielsen10,Pfeffer12,Malkin14,Meysman18}. To this end, the cable bacteria embed a network of thin fibers ($\sim$ 50 nm diameter) within their cell envelope\cite{Cornelissen18}. These fibers  are arranged in a highly regular, equidistant and parallel pattern, and essentially act as a power line network that connects all cells within the centimeter-long bacterial filaments.        

Intriguingly, these fibers from cable bacteria have been shown to exhibit electronic properties that are unprecedented in biology. Most prominently, electrical characterization revealed that the fibers display an extremely high conductivity (5-500 S cm$^{-1}$)\cite{Meysman19,Bonne20,Pankratov24,vanderVeen24}, which   substantially exceeds that of other known conductive biomaterials, like polymerized multi-heme cytochromes ($< 50 \times 10^{-3}$\,S\,cm$^{-1}$)\cite{Wang19,Wigginton07,Jiang20}. As such, the fibers produced by cable bacteria may provide a promising design principle for novel materials \cite{Bonne25} within the emerging field of protein 
bioelectronics \cite{Bostick18,Garg18,Futera23,Garg24}. Yet, the fundamental question of what biological electron transport mechanism is capable of sustaining such a high conductivity, remains presently unresolved.     

Detailed electrochemical characterization has shown that the fiber conductance in cable bacteria exhibits conspicuous properties, including the absence of redox activity and the lack of a gating response \cite{Pankratov24}. Likewise, the conductance shows a markedly weak thermal activation, with an activation energy of $\sim$40-50 meV at room temperature, which is 5 times lower than in multiheme cytochrome complexes\cite{vanderVeen25}. When the conduction is interpreted in terms of the conventional hopping picture, then very large effective transfer distances must be assumed (> 10 nm) for the ET to remain within the classical non-adiabatic Marcus regime \cite{vanderVeen24}. This transfer distance is an order of magnitude larger than typical nearest neighbor distances in conductive proteins and other organic molecular materials, and would suggest that charge carriers are (partially) delocalized \cite{vanderVeen24}. As such, it has been argued that the charge transport in the fibers may be reminiscent of that in high mobility molecular organic semiconductors\cite{Pankratov24,vanderVeen24}, where charge transport can occur via a transient delocalization mechanism with partially delocalized charge carriers\cite{Fratini16,Fratini17,Fratini20,Heck15,Xie20,Roosta22,Giannini19,Giannini20,Giannini22acr,Giannini23,Elsner24,Sneyd22}. In single crystals of rubrene, holes can delocalize over  multiple (>10) molecules at room temperature, which induces charge transfer over distances of 10 nm within the lattice\cite{Giannini20,Elsner24}. This length scale is hence similar to the one inferred for the transport in the conductive fibers of cable bacteria. 

To attain a better mechanistic understanding of the charge transport in fibers of cable bacteria, one requires atomistic molecular models that serve as a structural basis for computational simulations\cite{Polycarpou25}. Very recently, considerable progress was made on this topic, which we will take advantage of here. Based on high-resolution microscopy and spectroscopy, it was shown that each fiber centrally embeds an entangled bundle of ~10 nanoribbons\cite{Meysman26}. Each nanoribbon comprises a long, thin wire (1.4 nm diameter; > 100 nm length), which is composed of nickel bis(1,2-dithiolene) (NiBiD) oligomers as the molecular building block. In the model proposed, these planar molecules are axially aligned and slip-stacked into a multilayer, one-dimensional metal organic framework (MOF) that is a few molecular layers thick, thus forming the nanoribbon\cite{Meysman26}. The nanoribbons bear a strong resemblance to synthetic Ni-based coordination polymers, which exhibit comparably high electrical conductivities\cite{Xie22}. 

These proposed nanoribbons provide a very new type of supramolecular MOF structure. Accordingly, it is adamant to properly characterize the electronic properties. To this end, we performed Density Functional Theory (DFT) calculations of prospective nanoribbons as to verify whether such structures can truly support the efficient conductance that has been measured in current-voltage measurements. Our aim was to obtain further insight into the molecular structure of the nanoribbons, as currently, certain aspects are not precisely known, such as the number of Ni centers in the NiBiD oligomers as well as the packing structure of these NiBiD units. Accordingly, we evaluated possible atomistic configurations and potential packing structures of nanoribbons, and ranked their stability by calculation of their cohesive energy. We then calculated the relevant electron transfer parameters governing charge transport in these generated structures, i.e., reorganization energy ($\lambda$) and electronic coupling ($H_{\text{ab}}$) for both electron holes and excess electrons. These parameters provide a insight into the capability for efficient charge transport. More detailed first principles-based calculations of charge mobility and conductivity are beyond the scope of this letter and are left for future work. 

The nanoribbon structures evaluated were selected in accordance with the structural model as proposed in Ref.\cite{Meysman26}. The basic building block is the NiBiD unit shown in Fig.~\ref{fig:1}a, which is a nickel ethylene tetrathiolate (Ni$_3$(ett)$_4$H$_2$) compound containing three Ni centers terminated by two H atoms on opposite ends of the molecule. The electronic spin ground state of this molecule is a closed-shell singlet. While Raman spectroscopy indicates that multiple Ni centers are present \cite{Meysman26}, the actual value of $n$ (the number of Ni centers) in the NiBiD oligomer remains uncertain. We selected $n$=3 as a relevant, representative example, so that the molecule displays extended conjugation, but is small enough to keep the simulation effort tractable. Likewise, the end-capping of the NiBiD oligomer is currently not resolved \cite{Meysman26}; two terminal H atoms are intended to represent potential sites for disulfide linkages to the protein environment. Comparison of simple -H capping versus -S-CH$_3$ capping (which emulates cysteine capping as in Ref.~\cite{Meysman26}) indicated that the electronic structure is largely insensitive to the identity of the terminating group (see Fig.~\ref{fig:S1}).

Electronic structure calculations were carried out, unless stated otherwise, with a global hybrid functional based on PBE\cite{Perdew96} with 20\% GGA exchange replaced by Hartree-Fock exchange (HFX), including the D3 dispersion correction \cite{Grimme10} and Becke-Johnson damping \cite{Becke05}, denoted PBE20-D3(BJ). The fraction of exact exchange was chosen such that Koopmans' (Janak's) theorem \cite{Koopmans34,Janak78} for the ionization potential and electron affinity of Ni$_3$(ett)$_4$H$_2$ was obeyed (see Fig.~\ref{fig:koop}). However, calculations with the standard PBE0 (25\% HFX)\cite{Adamo99} functional gave similar results, suggesting that simulation results are robust against the choice of HFX fraction.

The optimization of nanoribbon packing structures required a large number of total energy and force calculations, which are computationally expensive with hybrid functionals. Thus, nanoribbon packing structures and cohesive energies were calculated using PBE-D3(BJ). This functional gave good performance on a benchmark set of small aliphatic and aromatic dimers with a mean absolute deviation ratio (MADR) compared to CCSD(T) of 9\% (similarly, PBE0-D3(BJ) MADR = 11\%) \cite{Tsuzuki20}. All calculations were carried out with the CP2K program package\cite{Kuhne20}. In the following we present calculations on monomer and dimer structures of Ni$_3$(ett)$_4$H$_2$ before presenting results for extended nanoribbon structures involving multiple stacked units.    
   
The LUMO and HOMO of Ni$_3$(ett)$_4$H$_2$ are shown in Fig.~\ref{fig:1}a-b, respectively. Both orbitals are primarily located on the ett ligands with very little or no contribution from the Ni d states. Notably, the LUMO is approximately C$_2$-symmetric about the long and short molecular axes and the axis perpendicular to the molecular plane. The orbitals are qualitatively very similar to those reported in Ref.~\cite{Meysman26} for the same oligomer with the B3LYP functional, and resemble those of related Ni-based dithiolene complexes as reported in Refs.~\cite{Kato04,Petrenko06}. The reorganization energy for charge transfer between two Ni$_3$(ett)$_4$H$_2$ molecules (at infinite separation) is 158 meV for electrons and 109 meV for holes, with the largest structural changes being the contraction of the central C-C bond and Ni-S bond of the dithiolene rings, respectively for electrons and holes (all <0.03~\AA). A smaller excess electron reorganization energy of 40 meV has been reported for Ni$_3$(ett)$_4$H$_2$ with the GGA functional BP86 \cite{Polycarpou25}. However, it is well known that GGA functionals tend to underestimate reorganization energy due to their deficiency in underestimating the energy for bond distortion \cite{Giannini19,McKenna12}. Replacing the terminal H atoms by S-CH$_3$, the reorganization energy changes only very little, 162 meV for electrons and 119 meV for holes. Overall, the predicted reorganization energies are typical of organic pi-conjugated molecules\cite{Giannini19}.   

Electronic couplings in Ni$_3$(ett)$_4$H$_2$ dimers for electrons and holes were calculated using the projection operator based diabatization (POD) method (available in the default CP2K package; see  Refs.\cite{Futera17,Ziogos21jcp1} for theoretical details). When dimers are positioned in perfect co-facial alignment, the electronic coupling decays exponentially with separation distance with an exponential distance decay constant of 2.9~\AA$^{-1}$ (non-centrosymmetric alignment) and 3.1~\AA$^{-1}$ (centrosymmetric alignment) for holes and 2.6~\AA$^{-1}$ in both alignments for electrons, Fig.~\ref{fig:2}a-b. These values are again typical for organic pi-conjugated molecules\cite{Kubas14jcp,Kubas15pccp,Ziogos21jcp1}. Here, (non-)centrosymmetric refers to absence or presence of inversion symmetry in the dimer pair due to the relative orientation of the H atoms on each monomer, as defined in Fig.~\ref{fig:1}c-d. Electronic coupling for electrons is the same for both H-atom orientations due to the approximate C$_2$ symmetry of the LUMO. 

Next we investigated the effect of displacing one monomer with respect to the other along the long and short molecular axes, respectively referred to as ``axial" and ``lateral" displacement. The axially and laterally displaced dimer configurations are labeled Ax$N$ and Lat$N$, respectively. The displacement index $N$ indicates the magnitude of the shift versus the reference cofacial alignment: $N\!=\!0, \dots,14$ in the axial direction (Fig~\ref{fig:1}e) and $N\!=\!0, \dots,4$ in the lateral direction (Fig.~\ref{fig:1}f). The electronic coupling of the axially displaced configuration exhibits a distinctive zig-zag pattern (Fig.~\ref{fig:2}c-d) which is particularly pronounced for holes where maxima occur when S atoms of the two NiBiD units are on top of one another (Fig. ~\ref{fig:2}c). In contrast, for laterally displaced monomers, the electronic coupling steadily decreases or increases with lateral index (Fig.~\ref{fig:2}e-f). All trends in electronic coupling can be readily explained in terms of constructive and destructive frontier orbital overlaps, which sensitively depend on the axial and lateral displacements; this is well established for other pi-conjugated dimers \cite{Coropceanu07}. Crucially, the dimer calculations show that in certain dimer configurations high electronic couplings are obtained, which exceed the critical threshold for charge delocalization, $H_{\text{ab}} > \lambda/2$ (indicated by a dashed line in~\ref{fig:2}c-f). These high electronic couplings persist even at relatively large axial and lateral displacements. Future microscopy and spectroscopic investigations should examine whether Ni$_3$(ett)$_4$H$_2$ nanoribbons preferentially pack into such electronically strongly coupled molecular configurations. Since the effective packing structure of nanoribbons remains unknown, we may take some clues from comparable Ni-bis(dithiolene) materials for which the packing structure has been experimentally determined \cite{Xie22}. These materials tend to form two packing motifs, denoted AA and AB. These motifs correspond to the direction of translation along the non-periodic lateral axis, which can either occur in the same direction (AA; Fig.~\ref{fig:1}g) or alternate between adjacent layers (AB; Fig.~\ref{fig:1}h). 

We constructed unit cells containing two Ni$_3$(ett)$_4$H$_2$ molecules in their centrosymmetric and non-centrosymmetric orientation. The molecular structure was determined from geometry optimization of the single molecule in vacuum using the PBE-D3(BJ) functional. In a first coarse scan of possible AA and AB packing structures, we generated axially and laterally displaced dimers at a fixed stacking distance corresponding to the potential energy minimum for the dimer in vacuum ($\approx3.55$ \AA, see Fig.~\ref{fig:S3}). The unit cell vectors were chosen such that the axial, lateral and stacking distances between monomers within and across the unit cell boundaries were equal. The unit cell was then periodically replicated in the axial and in the stacking direction, but not in the lateral direction. According to the spectroscopic data, nanoribbons consist of 2 to 5 molecular layers in the stacking direction \cite{Meysman26}, and hence, they are finite along the stacking direction. Still, we also applied periodic boundary conditions along the stacking direction to avoid artificial surface effects that would otherwise be introduced. Given that we chose displacements between all neighboring monomers to be equal, axial displacements smaller than axial index 8 are inherently energetically unfavorable due to steric overlap between molecules of adjacent cells (see Fig.\ref{fig:ee} and Table \ref{tbl:ee}). To quantify the relative stability of each packing structure, we calculated the cohesive energy, $E_{\text{coh}}$ (defined positively), using the PBE-D3(BJ) functional,
 \begin{equation}\label{cohesive}
     E_{\text{coh}} = - ( E_{\text{tot}} - N E_{\text{mono}} )
 \end{equation}
where $E_{\text{mono}}$ is the energy of the monomer in vacuum, $E_{\text{tot}}$ is the total energy of a unit cell and $N$ is the number of monomers in the unit cell ($N\!=\!2$ unless stated otherwise). The greater the cohesive energy, the more energetically stable and thus likely a given configuration is.

\begin{figure}[h!]
    \centering
    \includegraphics[width=1\linewidth]{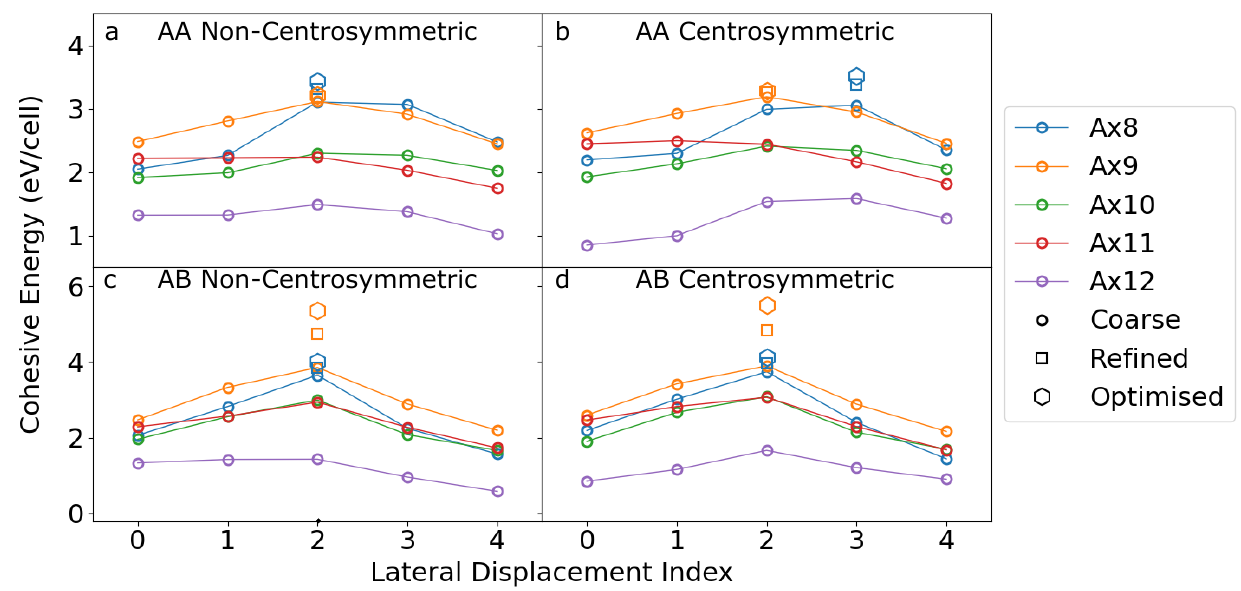} 
    \caption{\textbf{Packing stability for 2D NiBiD nanoribbon model structures.} Cohesive energies per unit cell, calculated using PBE-D3(BJ), are shown for 2D nanoribbons with different packing motifs (AA and AB) and dimer inversion symmetry configurations (centrosymmetric and non-centrosymmetric): \textbf{a} AA non-centrosymmetric, \textbf{b} AA centrosymmetric, \textbf{c} AB non-centrosymmetric, and \textbf{d} AB centrosymmetric. Coarse points (circles) correspond to structures generated across axial displacement indexes Ax8-Ax12, indicated by color, and lateral displacement indexes Lat0-Lat4, shown on the x axis. `Refined' points (squares) and `Optimized' points (hexagons) correspond to structures generated by a refinement procedure and subsequent geometry optimization of the refined structure, respectively. These points are plotted using the same indexing scheme (color and x-axis) as the coarse structure from which refinement was initiated. See Fig.\ref{fig:1}e-f for definition of displacement index.}
    \label{fig:3}
\end{figure}

Fig.~\ref{fig:3} displays the cohesive energies for AA and AB packed structures as a function of the lateral displacement (Lat0-Lat4) for different axial displacements (Ax8-Ax12) at a constant stacking distance. The cohesive energy is governed by the balance between dispersion interactions and Pauli repulsion. In both the AA and AB packed structures, the configurations with the smallest axial displacements (Ax8 and Ax9), have the greatest dispersion interaction and cohesive energy for all lateral displacements investigated. Furthermore, a preference for lateral displacements Lat2 and Lat3 is observed, which can be attributed to reduced orbital overlap and, therefore, decreased Pauli repulsion, despite slightly weaker dispersion interactions. In addition, AB packing structures are generally more stable than their AA counterparts, owing to the shorter interlayer next-nearest-neighbor distance in AB, which provides additional dispersion stabilization (see Fig.~\ref{fig:AAvsABcoh}). Overall, this coarse scan identified the Ax8 and Ax9 packed structures as the
most stable configurations across both AA and AB packing motifs in both centrosymmetric and non-centrosymmetric configurations.

In a subsequent step, we refined the AA and AB packings and the corresponding unit cell dimensions starting from configurations with the greatest cohesive energy in the Ax8 and Ax9 series (Lat2 or Lat3) by varying the axial and lateral displacements and the stacking distance in increments of 0.01~\AA. Despite refined structures no longer aligning with the initial axial and lateral displacement indexes, we continue to refer to them with the indexes of the structure from which each refinement was initialized (e.g. refined Ax8). AB motifs remained more stable than AA motifs upon refinement. Yet, the magnitude of the increase in cohesive energy varied across configurations, with modest stabilization upon refinement for AA Ax8, AA Ax9, and AB Ax8 initial structures, but a marked increase in cohesive energy for AB Ax9 initial structure (Fig.~\ref{fig:3}, data points indicated by squares). Analysis of the refined AB Ax9 configuration indicated that the large increase in cohesive energy originated from alignment of Ni and S atoms between layers, accompanied by a significant reduction in interlayer separation from $\approx 3.5$~\AA~to $3.0$~\AA, enabling the formation of a weak axial Ni-S coordination bond. Similar Ni-S interactions have been highlighted in other Ni-based dithiolene complexes \cite{Xie22}. Such interlayer contacts are geometrically not possible in Ax8 configurations. Interestingly, they would be possible in AA Ax9, but are energetically not favorable in this configuration. 

\begin{figure}[h!]
    \centering
    \includegraphics[width=1\linewidth]{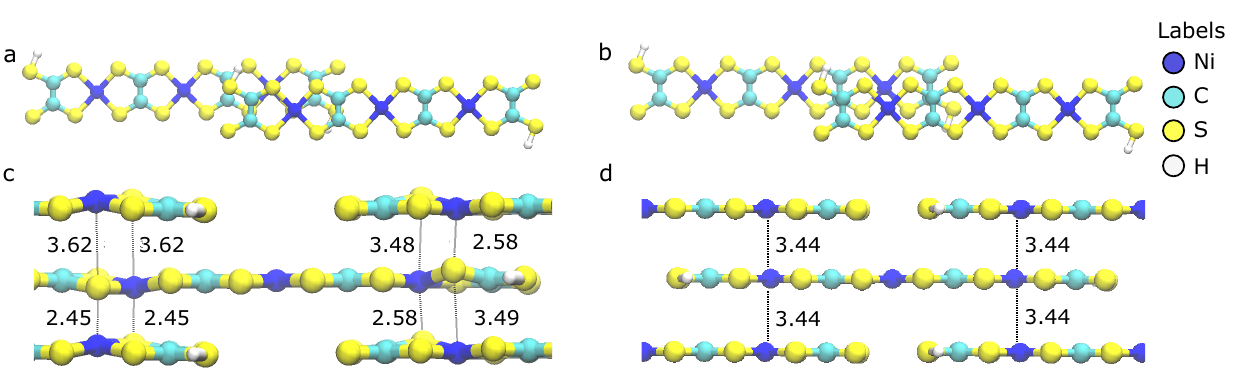} 
    \caption{\textbf{Structural comparison of optimized centrosymmetric AB Ax9 and AB Ax8 nanoribbon structures.} Dimer pairs extracted from optimized 2D nanoribbon sheets are shown from above for centrosymmetric AB Ax9 (\textbf{a}) and AB Ax8 (\textbf{b}). The optimized AB Ax9 structure shows clear interlayer alignment of Ni and S atoms; this alignment is not observed in the optimized AB Ax8 structure. Snapshots of the corresponding 2D nanoribbon are shown in \textbf{c} and \textbf{d} for optimized AB Ax9 and AB Ax8, respectively. In \textbf{c}, the Ni and S atoms distort out of their $\text{NiS}_4$ square planes, producing shorter and longer interlayer Ni-S distances whereas the shorter distances correspond to interlayer Ni-S bonds. This reduction in interlayer distance is not observed in \textbf{d}, where the individual $\text{Ni}_3\text{(ett)}_4\text{H}_2$ units remain undistorted and planar.}
    \label{fig:4}
\end{figure}

Final "optimized" structures were obtained by geometry optimization of the refined structures under fixed unit cell parameters. AB Ax9 underwent marked structural relaxations along with large energetic stabilization (Fig.~\ref{fig:3}c-d, hexagons in orange), whereas the other structures exhibited minimal relaxations and energetic stabilization (AA Ax8, AB Ax8 blue hexagons Fig.~\ref{fig:3}a-d; AA Ax9 orange hexagons Fig.~\ref{fig:3}a-b). In AB Ax9, Ni and S atoms involved in close interlayer contacts distorted out of their NiS$_4$ square planes upon optimization, giving rise to shortened interlayer Ni-S distances ($\approx 2.5$~\AA) with the lower plane, resulting in the formation of a 5-fold coordinated Ni center, and longer interlayer Ni-S distances with the upper plane ($\approx 3.5$~\AA) (see Fig.~\ref{fig:4}c). Each monomer can form 4 such interactions (2 at each end) with neighboring monomers in the layers above or below. Conspicuously, in the optimized AB Ax9 stacking, the terminal ends of the symmetric NiBiD molecules behaved differently. The right-hand side formed interlayer coordination interactions with both layers above and below, whereas the left-hand side formed interactions only with the layer below. This resulted in symmetry breaking between monomers and the emergence of non-equivalent sites. Across all packing structures analyzed, AB Ax9 packing emerged as the most stable configuration, owing to its unique ability to form interlayer Ni-S coordination through structural reorganization (Fig.~\ref{fig:4}a,c), with AB Ax8 being the second most stable structure (Fig.~\ref{fig:4}b,d).

While the cohesive energy determines the packing structure of the NiBiD molecule, the electronic coupling governs its charge transport capability. Thus, it is of interest to evaluate the electronic coupling for the different packing structures generated. In the standard approach for calculating electronic couplings in the condensed phase, one extracts molecular dimers from the bulk and carries out coupling calculations on them in vacuum. We found that this approach was not applicable for Ni$_3$(ett)$_4$H$_2$ because the diabatic frontier orbitals of the dimer exhibited strong polarization with respect to the frontier orbitals in the single monomer in vacuum (Fig.\ref{fig:podprob}a,c). This polarization does not happen in the periodic environment of a nanoribbon, where the diabatic frontier orbitals resemble closely the frontier orbital of the single monomer in vacuum (Fig.\ref{fig:podprob}b,d). Thus we computed electronic couplings between neighboring molecules directly in the condensed phase using periodic boundary conditions exploiting the unique capabilities of the POD method (see Figs.~\ref{fig:podprob} and \ref{fig:primvsext} and Tables \ref{vac:con} and \ref{tbl:LR}). 

For both hole and excess electron coupling, Ax8 configurations (Fig.~\ref{fig:5}, data in blue) exhibited significantly larger couplings than Ax9 configurations (Fig.~\ref{fig:5}, data in orange), consistent with the frontier orbital overlap arguments used above to explain the trends for dimers in vacuum (Fig.~\ref{fig:2}). For most configurations, refinement and geometry optimization had little effect on the coupling values as the structures did not deviate significantly from their initial geometries (Fig.~\ref{fig:5}, data indicated by squares and hexagons, respectively). As with cohesive energy, AB Ax9 was the outlier. A marked increase in electronic couplings was observed upon refinement (Fig.~\ref{fig:5}b,d, squares in orange) which was driven by the significantly reduced stacking distance (axial Ni-S distances $\approx 3.0$~\AA). In the optimized structure, the formation of short and long axial Ni-S distances ($\approx 2.5$~\AA~and $\approx 3.5$~\AA, Fig.~\ref{fig:4}c), resulted in 4 non-equivalent couplings, one with a very large value corresponding to the dimer pair with two short axial Ni-S bonds and three with smaller values (Fig.~\ref{fig:5}b,d, hexagons in orange). While the precise pattern of short and long axial Ni-S distances may depend on the thickness of the nanoribbon, they will always result in the emergence of non-equivalent electronic couplings. 

Efficient charge transport in the transient delocalization regime occurs when the electronic coupling exceeds $\lambda/2$ (Fig.~\ref{fig:5}, dashed lines). Using this criterion, a clear contrast emerges between AB Ax8 and AB Ax9 configurations. In the AB Ax8 configuration each molecule is connected with its neighbors by couplings that exceed this threshold, for both excess
electron and hole. In this way, continuous strong coupling paths exist that support efficient delocalized transport. By contrast, in the AB Ax9 configuration only one molecular pair is strongly coupled with a value exceeding  $\lambda/2$  whereas others are weakly connected with values below $\lambda/2$. Thus, this structure does not form a continuous path of strong electronic couplings exceeding $\lambda/2$ - yet delocalized transport may still occur depending on the extent of eigenstate delocalization and thermal electronic disorder. Investigation of such effects was however beyond the scope of this work and is left for future investigations.

Several previous experimental and computational studies of NiBiD-related complexes have provided detailed insight into molecular electronic structure and packing motifs \cite{Xie22,Petrenko06,Amb16,Kato04,Ray05}. However, simultaneous consideration of packing stability and intermolecular electronic coupling has not been carried out, to our best knowledge. Here, a clear trade-off between structural stability and charge transport is revealed within the NiBiD nanoribbons investigated. While the AB Ax9 configuration is the most stable structure due to its ability to form interlayer Ni-S coordination interactions, this same structural reorganization leads to asymmetric environments and non-equivalent couplings that may limit charge transport. In contrast, Ax8 configurations are less energetically stabilized, but maintain configurations that preserve a good orbital overlap between neighboring molecules. This gives rise to a continuous network of high electronic couplings exceeding $\lambda/2$ and enabling charge delocalization. Yet, it is important to note that in our idealized simulations, we considered only nanoribbons consisting of Ni$_3$(ett)$_4$H$_2$ oligomers in isolation, and hence we did not account for possible interactions with the surrounding environment, such as a protein scaffold, which is proposed to encapsulate the nanoribbons in the periplasm of cable bacteria \cite{Boschker21}. These environmental effects could tip the energetic balance between the different packing motifs and might stabilize configurations with favorable continuous electronic coupling networks that are not the most stable in isolation, such as the AB Ax8 structures investigated here.
 
In conclusion, we have evaluated a possible set of atomistic model structures for cable bacterial nanoribbons made of NiBiD oligomers. Some of these configurations support efficient delocalized electron transport beyond nearest-neighbor hopping, which could help explain the high conductivity and other unusual electrical properties (no redox activity, weak gating response, weak thermal activation) that are reported for these intriguing biological structures \cite{Meysman19,Bonne21,Pankratov24,vanderVeen24}. Future work should concentrate on the further experimental elucidation of the atomistic structure of the nanoribbons, which could provide a validation of the exploratory simulation results presented here. The current work has laid the foundation for state-of-the-art non-adiabatic dynamics simulations \cite{Giannini19,Giannini20,Giannini22acr,Giannini23,Elsner24} of NiBiD oligomer-based nanoribbons, which will quantitatively elucidate the intrinsic charge mobilities and transport mechanism. 

\section{Associated Content}
Supporting Information giving full computational details on the preparation of nanoribbons, their coordinates in xyz format 
and the calculation of charge transfer properties.  

\section{Acknowledgments}
O.R. was supported by a PhD studentship provided by the LCN CDT for quantum technologies, EPSRC Grant number EP/W524335/1. This work used the Archer2 UK National Supercomputing Service via our membership in the UK’s HEC Materials Chemistry Consortium, which is funded by EPSRC (Grants EP/L000202 and EP/R029431), as well as the UK Materials and Molecular Modeling (MMM) Hub, which is partially funded by EPSRC (Grant EP/P020194). FJRM was supported by the Volkswagen Stiftung (Quantum Biology program, project 0200347-01) and the Research Foundation Flanders (Fonds Wetenschappelijk
Onderzoek), grants S004523N and G0ADR25N.

\nocite{Elmaslmane18,Bondi64}

\bibliography{bibliography}

\begin{figure}[p]
    \centering
    \includegraphics[width=1\linewidth]{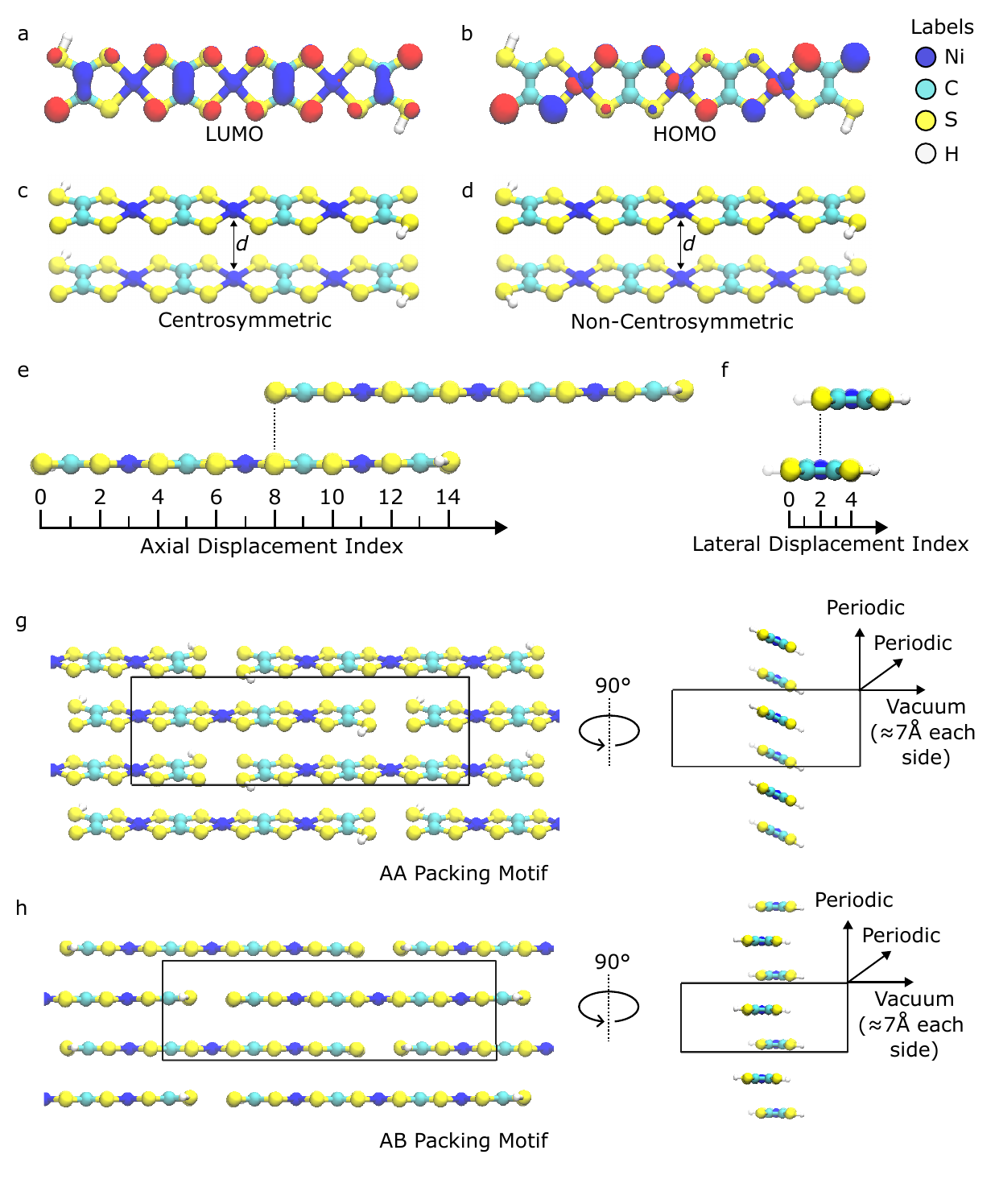}
\end{figure}

\clearpage

\captionof{figure}{\textbf{Frontier orbitals of the NiBiD monomer and structural parameters defining dimer and nanoribbon packing structures.} \textbf{a} and \textbf{b} show the LUMO and HOMO of the monomer, respectively, plotted at an isosurface value of $\pm 0.03$, with red and blue denoting opposite orbital phases. \textbf{c} and \textbf{d} define centrosymmetric and non-centrosymmetric configurations, respectively. Centrosymmetry refers to the presence or absence of inversion symmetry in a dimer pair, best indicated by the relative orientation of terminal H atoms. The interlayer stacking distance, $d$, is also indicated. Indexing schemes for axial and lateral displacements are shown in \textbf{e} and \textbf{f}, respectively. Each index corresponds to translating one monomer along the axial or lateral axis such that its terminal S atom aligns above the indexed atom, $N$, of the other monomer. These configurations are denoted Ax$N$ and Lat$N$, as illustrated for Ax8 in \textbf{e} and Lat2 in \textbf{f}. AA (\textbf{g}) and AB (\textbf{h}) packing motifs are shown from two viewing orientations, with the primitive unit cells indicating the periodic and vacuum directions. In AA packing, each layer is laterally shifted in the same direction relative to the layer below. In AB packing, the direction of this lateral shift alternates between layers.}
\label{fig:1}

\clearpage

\clearpage

\begin{figure}[h!]
    \centering
    \includegraphics[height=1\linewidth]{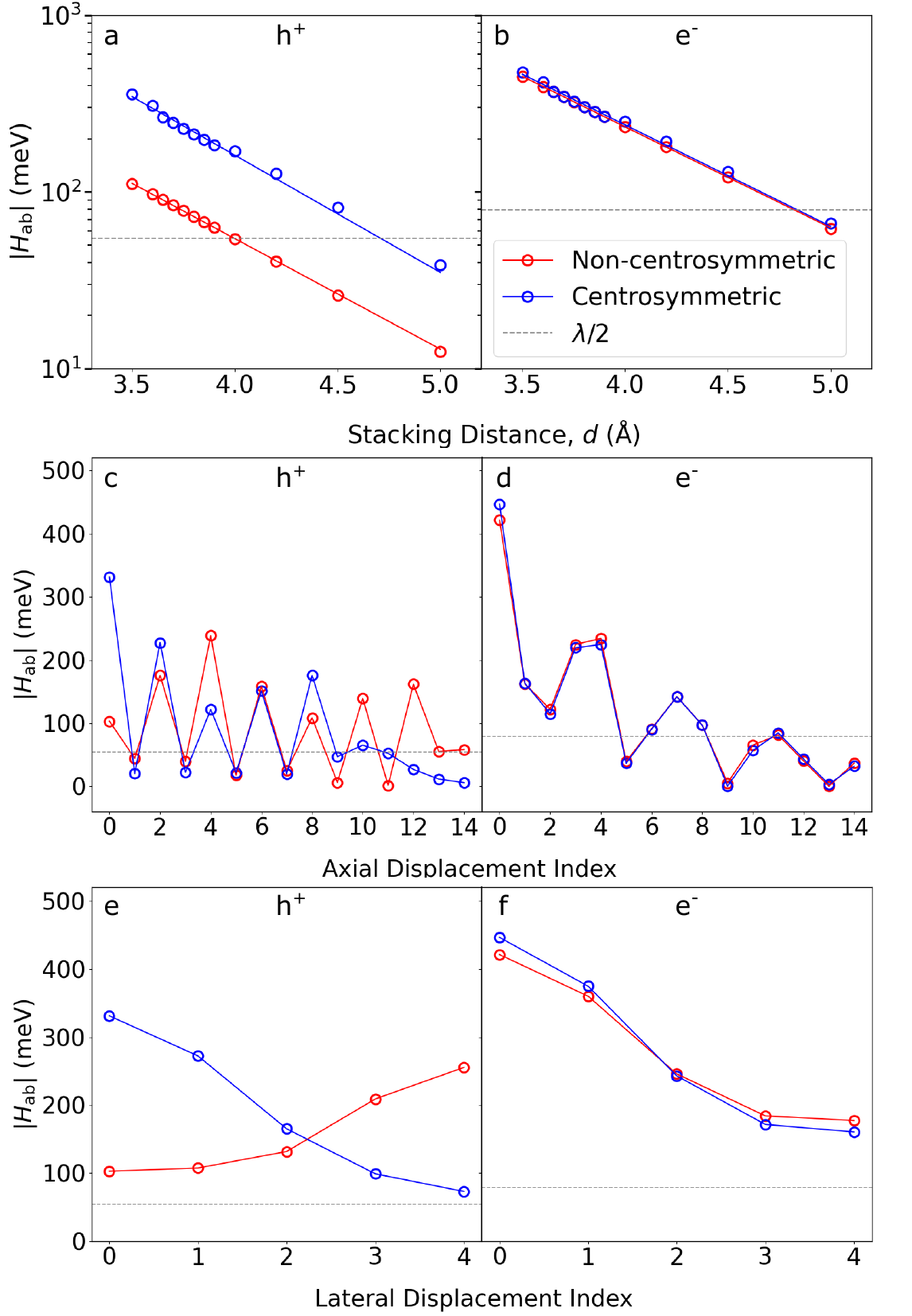} 
    \caption{\textbf{Electronic coupling of NiBiD dimers in vacuum in centrosymmetric and non-centrosymmetric configurations.} Electronic coupling for hole (HOMO-HOMO, \textbf{a}) and excess electron (LUMO-LUMO, \textbf{b}) for the fully cofacial dimer pair (Ax0Lat0) is shown as a function of interlayer stacking distance, $d$. The distance dependence is fitted to $|H_{\text{ab}}| = Ae^{-\beta d/2}$, giving decay constants of 2.9 and 3.1~\AA$^{-1}$ for hole coupling in the non-centrosymmetric and centrosymmetric configurations, respectively, and 2.6~\AA$^{-1}$ for excess-electron coupling in both configurations. Hole (\textbf{c}) and excess electron (\textbf{d}) coupling are plotted against axial displacement index at fixed interlayer stacking distance (3.55~\AA) and lateral displacement (Lat0). Similarly, hole (\textbf{e}) and excess electron (\textbf{f}) coupling are plotted against lateral displacement index at fixed interlayer stacking distance (3.55~\AA) and axial displacement (Ax0, see Fig.\ref{fig:1}e-f for definition of displacment index). The threshold $\lambda/2$, where $\lambda$ is the reorganization energy for hole or excess electron transport, is also shown; coupling values above this threshold are typically associated with efficient charge transport via the transient delocalization mechanism.} 
    \label{fig:2}
\end{figure}
\clearpage

\clearpage

\clearpage

\begin{figure}[h!]
    \centering
    \includegraphics[width=1\linewidth]{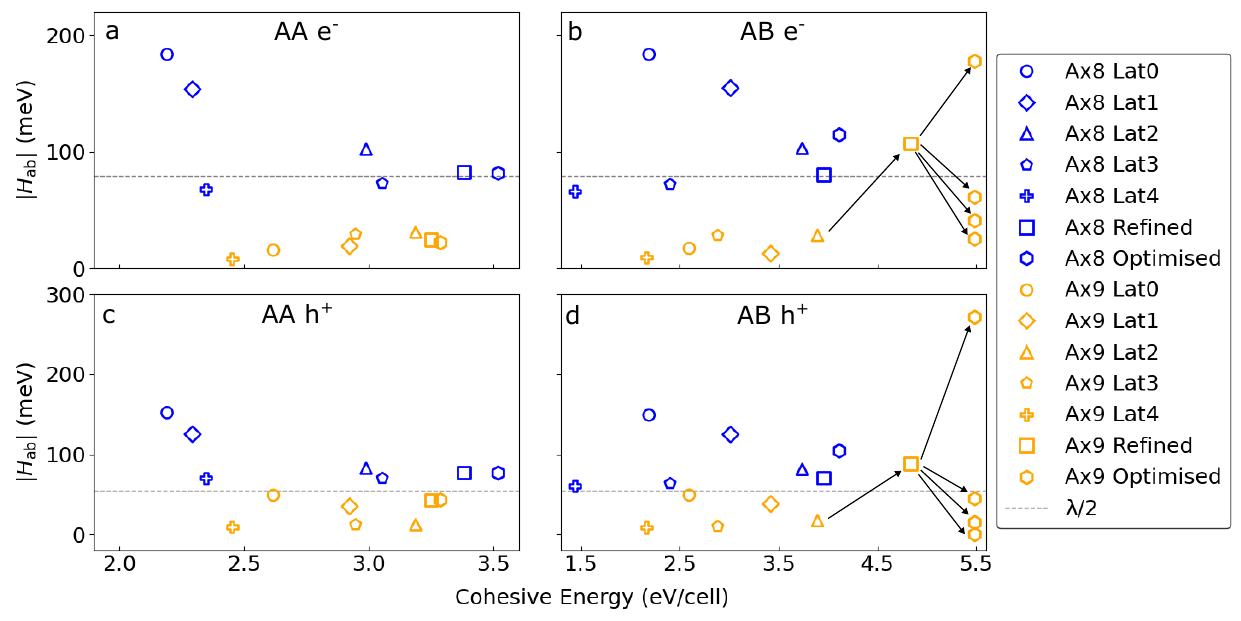} 
    \caption{\textbf{Electronic coupling versus cohesive energy for centrosymmetric nanoribbon model structures.} Electronic coupling (y axis) and cohesive energy (x axis) are plotted for centrosymmetric packing structures, with color indicating axial displacement index and marker shape indicating lateral displacement index. Refined and optimized structures are shown as squares and hexagons, respectively. AA and AB packing motifs are shown for excess electron coupling in \textbf{a} and \textbf{b}, respectively, and for hole coupling in \textbf{c} and \textbf{d}, respectively. Dashed gray lines indicate the $\lambda/2$ threshold, as in Fig.~\ref{fig:2}. Black arrows in \textbf{b} and \textbf{d} indicate the progression from 'Coarse' to 'Refined' structures (triangle to square) and from 'Refined' to 'Optimized' structures (square to hexagon) for the highest stability structures. Analogous data for non-centrosymmetric configurations are shown in Fig.~\ref{fig:noncentro1d}.}
    \label{fig:5}
\end{figure}

\clearpage
\section*{Supporting information}
\renewcommand{\thefigure}{S\arabic{figure}}
\renewcommand{\thetable}{S\arabic{table}}
\renewcommand{\theequation}{S\arabic{equation}}
\setcounter{figure}{0}
\setcounter{table}{0}
\setcounter{equation}{0}

\begin{figure}[h!]
    \centering
    \includegraphics[width=1\linewidth]{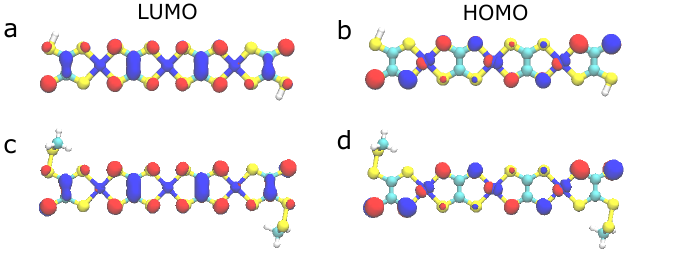} 
    \caption{\textbf{Frontier orbitals of (Ni$_3$(ett)$_4$H$_2$)R$_2$ with -H and -SCH$_3$ terminating groups (R).} \textbf{a} and \textbf{b} show the LUMO and HOMO, respectively, for R = -H; \textbf{c} and \textbf{d} show the LUMO and HOMO, respectively, for R = -SCH$_3$. Red regions correspond to an isosurface value of -0.03; blue regions correspond to +0.03. The LUMO and HOMO distributions are qualitatively insensitive to the identity of the terminating group.}    
    \label{fig:S1}
\end{figure}


\clearpage
\subsection*{Optimization of Hartree-Fock exchange fraction}

\begin{figure}[h!]
    \centering
    \includegraphics[width=1\linewidth]{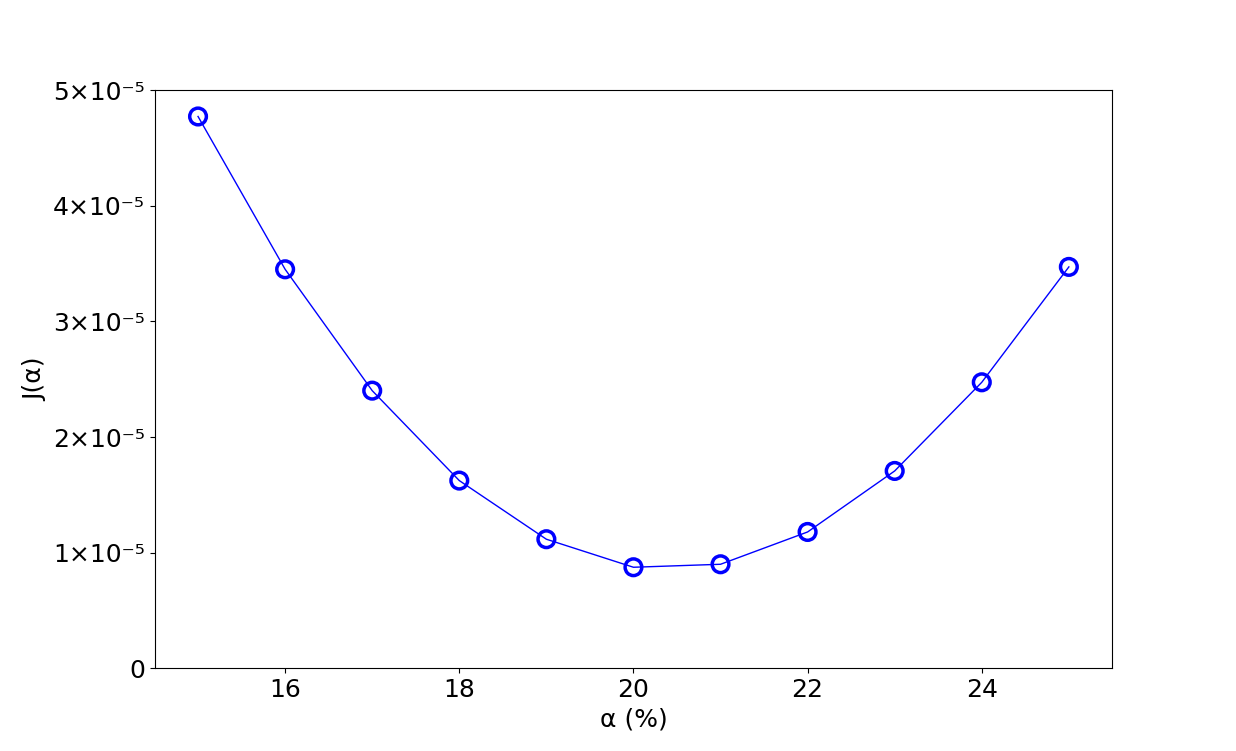} 
    \caption{\textbf{Optimal Hartree-Fock exchange fraction.} Using Koopmans' theorem, the square deviation between ionization energy (electron affinity) and orbital energy, $J(\alpha)$, is shown as a function of the HF exchange fraction, $\alpha$. A minimum in $J(\alpha)$ is found at $\alpha = 20 \% $.}
    \label{fig:koop}
\end{figure}

As outlined in Ref.~\cite{Elmaslmane18}, the fraction of Hartree-Fock exchange (HFX) in a hybrid functional can be tuned by enforcing Koopmans' (Janak's) theorem\cite{Janak78,Koopmans34}:
\begin{align}
I &= -\varepsilon^{0}_{\mathrm{HOMO}} \\
A &= -\varepsilon^{-}_{\mathrm{HOMO}}
\end{align}
where $I$ and $A$ are the vertical ionization energy and electron affinity, respectively. Here, $\varepsilon^{0}_{\mathrm{HOMO}}$ is the HOMO energy of the neutral monomer and $\varepsilon^{-}_{\mathrm{HOMO}}$ is the majority-spin HOMO energy of the anion. The corresponding quantities are given by:
\begin{align}
I(\alpha) &= E^{+}(\alpha) -E^{0}(\alpha)  \\
A(\alpha) &= E^{0}(\alpha) - E^{-}(\alpha) 
\end{align}
where $E^{+}(\alpha)$, $E^{0}(\alpha)$ and $E^{-}(\alpha)$ are the total energies of the cation, neutral and anion, respectively, with $\alpha$ referring to the percentage of HFX and each evaluated in the neutral geometry. The optimal fraction of HFX was determined by minimizing the square difference,$J(\alpha)$ , between orbital energies and the corresponding ionization energies and electron affinities, thereby enforcing Koopmans' (Janak's) condition. 
\begin{align}
DI(\alpha) &= E^{+}(\alpha) -E^{0}(\alpha) + \varepsilon^{0}_{\mathrm{HOMO}}(\alpha)  \\
DA(\alpha) &= E^{0}(\alpha) - E^{-}(\alpha) + \varepsilon^{-}_{\mathrm{HOMO}}(\alpha)
\end{align}
\begin{equation}
    J(\alpha) = DI^2(\alpha) + DA^2(\alpha)
\end{equation}
As shown in Figure \ref{fig:koop}, the optimal fraction of HFX for a single NiBiD molecule in vacuum was found to be 20\%. 

\clearpage

\subsection*{Determination of interlayer separation distance }
\begin{figure}[h!]
    \centering
    \includegraphics[width=1\linewidth]{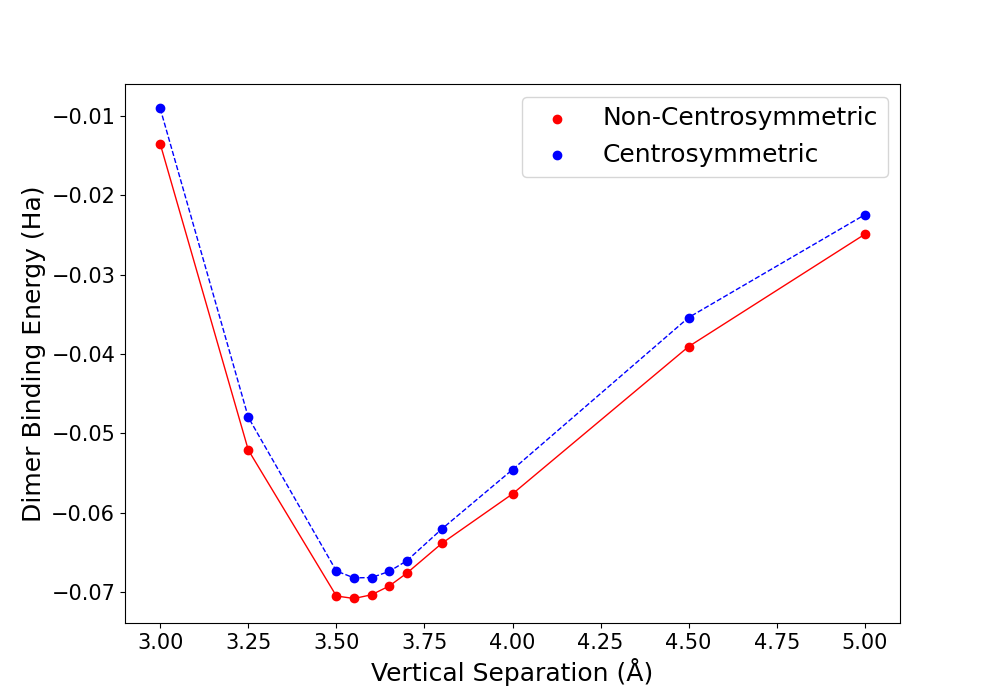} 
    \caption{\textbf{Interlayer stacking distance.} The dimer binding energy for fully cofacial dimer pairs is shown as a function of interlayer stacking distance. For both centrosymmetric and non-centrosymmetric configurations, minima are found at $\approx 3.55$~\AA. }
    \label{fig:S3}
\end{figure}
In order to compare axial and lateral displacement, it is necessary to first establish a constant interlayer separation distance. By taking the perfect co-facial dimer pair, we vary the interlayer separation distance (as in Figures \ref{fig:2}a and \ref{fig:2}b) and evaluate the dimer binding energy, $E_{\text{dim}}$ (defined negatively).
\begin{equation}
    E_{\text{dim}} = E_{\text{tot}} - 2E_{\text{mono}}
\end{equation}

This was computed using the PBE-D3(BJ) functional, consistent with cohesive energy calculations, and is shown in Figure \ref{fig:S3}. Both centrosymmetric and non-centrosymmetric orientations present a minima at $\approx 3.55$~\AA, hence for all calculations, unless otherwise stated (refined and optimized), an interlayer separation of $3.55$~\AA{}  was used.

\clearpage
\subsection{Unit cell construction and constraints}

\begin{figure}[h!]
    \centering
    \includegraphics[width=1\linewidth]{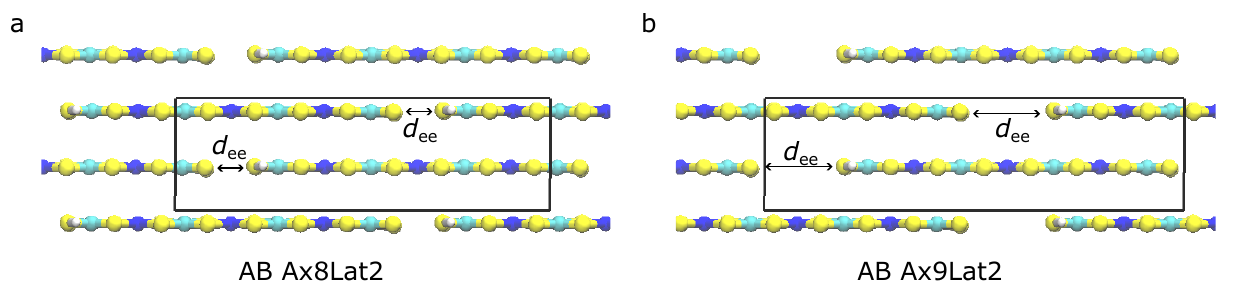} 
    \caption{\textbf{Intralayer NiBiD unit distance.} Intralayer distances, $d_{\text{ee}}$, between the ends of NiBiD molecular units are shown for packing structures with axial translations to axial index 8 (\textbf{a}) and axial index 9 (\textbf{b}).}
    \label{fig:ee}
\end{figure}
The intralayer NiBiD end-end distance, $d_{\text{ee}}$, is introduced  in Fig.\ref{fig:ee}. The end-end distance is controlled by axial displacement, with larger axial displacements giving larger end-end distances (see Ax8 and Ax9 in Fig.\ref{fig:ee}). This follows from the imposed condition of equal displacement between neighboring monomers within and across the unit cell boundaries. The intralayer end-end distance, $d_\text{ee}$ is given by:
\begin{equation}
    d_{\text{ee}} = 2d_{\text{ax}} - L_{\text{mono}}
\end{equation}
where $d_{\text{ax}}$ is the axial displacement and $L_{\text{mono}}$ is the length of an individual NiBiD unit in the axial direction (21.036 \AA). The resulting intralayer end-end distances are outlined for different axial displacement indexes in Table \ref{tbl:ee}. 
\begin{table}
\begin{tabular}{ccc}
\hline
Axial Index &  $d_{\text{ax}}$ (\AA) & $d_{\text{ee}}$ (\AA)\\
\hline
7 & 10.559 & 0.082 \\
8 & 12.075 & 3.114 \\
9 &  13.561 & 6.086\\
10 & 15.051 & 9.066\\
11 & 16.562 & 12.088\\
12 & 18.054 & 15.072\\
\hline
\end{tabular}
\caption{Intralayer end-end distances in packing structures constructed with different axial displacements.}
\label{tbl:ee}
\end{table}

This rationalizes the choice of a minimum axial displacement index of 8. Smaller axial displacement indexes give very short (or negative) intralayer end-end distances. For example, axial index 7 gives $d_{\text{ee}} = 0.082$ \AA~which is far below the sum of two sulfur van der Waals radii ($2r_{\text{vdW}}(\text{S}) = 3.60 $ \AA)\cite{Bondi64}. Packing structures generated with indexes below 8 are therefore expected to involve severe steric overlap and are thus highly energetically unfavorable.

\clearpage
\subsection*{AA vs AB Cohesive Energy}

\begin{figure}[h!]
    \centering
    \includegraphics[width=1\linewidth]{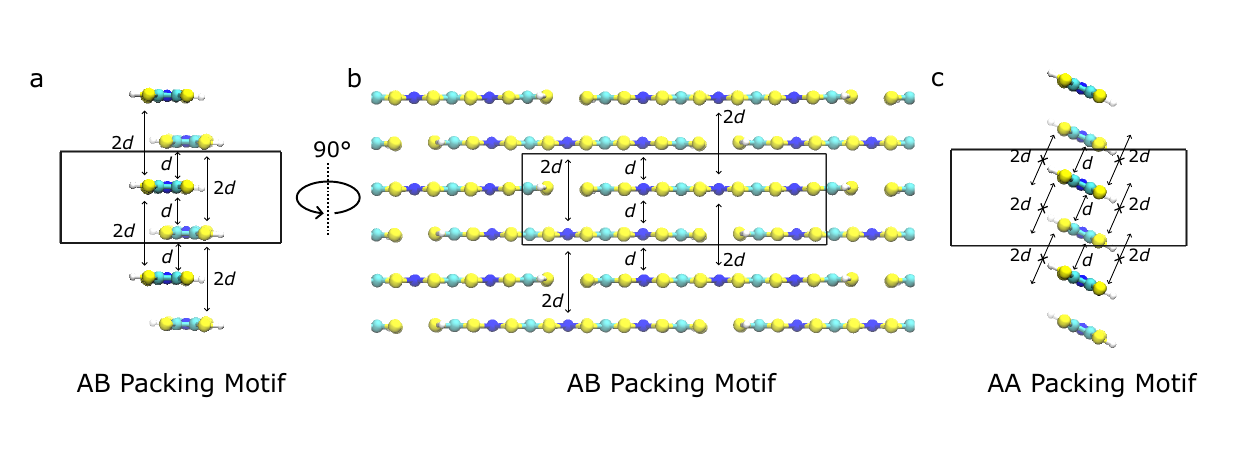} 
    \caption{\textbf{Next-nearest-neighbor interlayer interactions.} Possible nearest-neighbor ($d$) and next-nearest-neighbor ($2d$) interlayer interactions are indicated with double-headed arrows for the AB packing motif viewed along the lateral axis (\textbf{a}) and axial axis (\textbf{b}). Nearest-neighbor interactions ($d$) are similarly shown for the AA packing motif viewed along the lateral axis (\textbf{c}). The absence of next-nearest-neighbor interactions ($2d$) is indicated by crossed double-headed arrows.}
    
    \label{fig:AAvsABcoh}
\end{figure}

\clearpage

\subsection*{Frontier Orbitals from POD }

\begin{figure}[h!]
    \centering
    \includegraphics[height=1\linewidth]{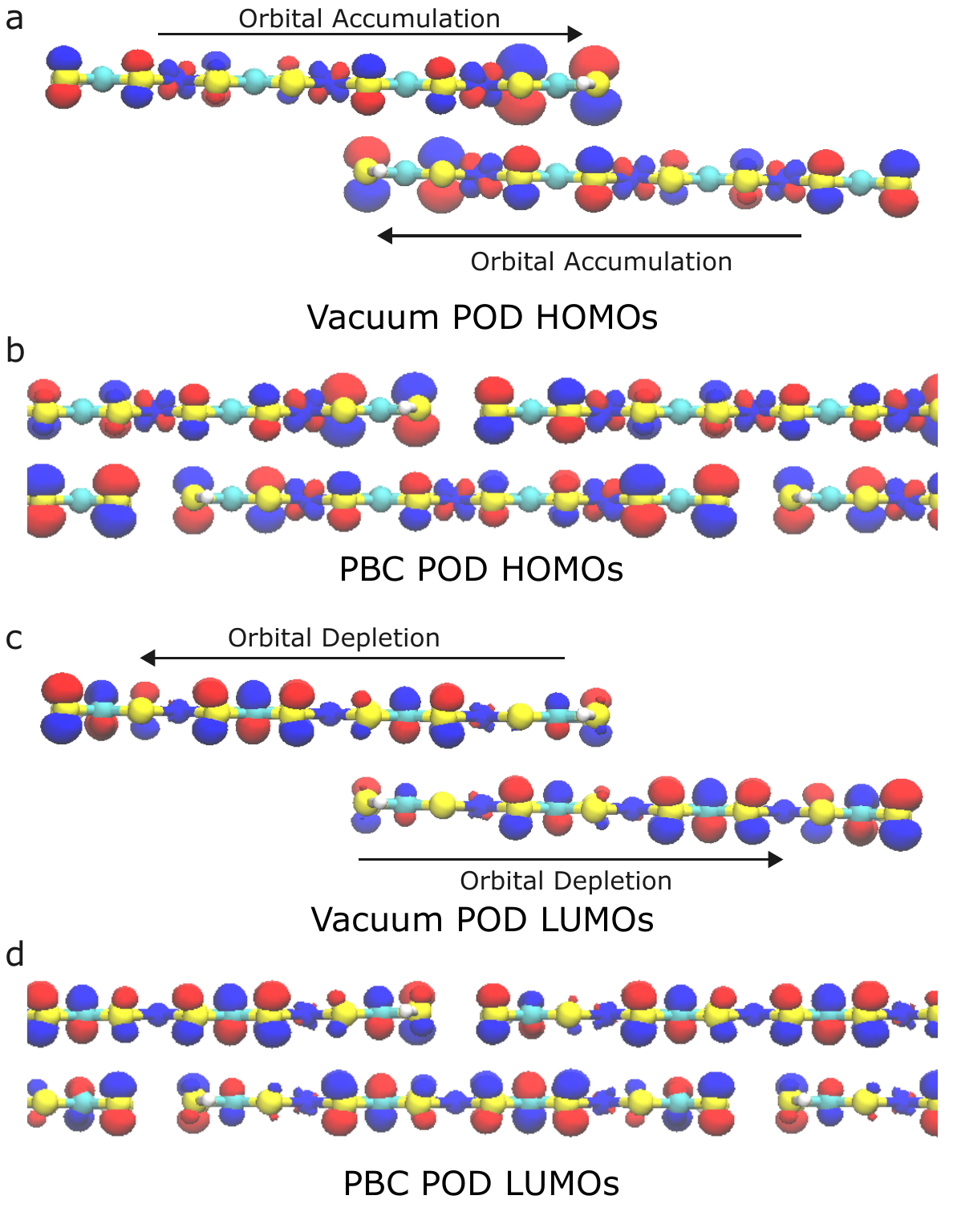} 
    \caption{ Frontier orbitals obtained from diabatization using the POD method. In \textbf{a} and \textbf{c} calculations are carried out for a NiBiD dimer in vacuum and the isosurfaces depict the HOMO (\textbf{a}) or LUMO (\textbf{c}) localized on each of the two  monomers forming the dimer as obtained from projection operator-based diabatization (POD). Notice the accumulation (\textbf{a}) or depletion (\textbf{c})  of orbital amplitudes in the molecular overlap region (i.e. orbital polarization). In \textbf{b} and \textbf{d} calculations are carried our for NiBiD nanoribbons in PBC and the isosurfaces depict again the POD HOMO (\textbf{b}) or POD LUMO (\textbf{d}) localized on each molecule of the periodic system. Notice that the localized frontier orbitals in the periodic system show little polarization and resemble closely the frontier orbitals of the monomer in the gas phase, Fig.\ref{fig:1}a-b, main text. }    
    \label{fig:podprob}
\end{figure}


\begin{table}
\begin{tabular}{ccccc}
\hline
Configuration & Vac ($\text{h}^{+}$) & PBC ($\text{h}^{+}$) & Vac ($\text{e}^{-}$) & PBC ($\text{e}^{-}$) \\
\hline
AB CS Ax8 Lat0 & 187 & 149 & 98.5& 184\\
AB CS Ax8 Lat1 & 154& 125 & 93& 155\\
AB CS Ax8 Lat2 & 85.8& 81.3 & 77& 103\\
AB CS Ax8 Lat3 & 33.3& 63.7& 64.7& 72.1\\
AB CS Ax8 Lat4 & 2.64& 60.7& 62.7& 66.1\\
\hline
AB NCS Ax8 Lat0 & 104 & 98.4 & 98.0 & 181 \\
AB NCS Ax8 Lat1 & 94.0 & 85.5 & 90.1 & 153.5 \\
AB NCS Ax8 Lat2 & 77.6 & 65.9 & 72.7 & 101.5 \\
AB NCS Ax8 Lat3 & 88.1 & 69.9 & 58.5 & 70.6 \\
AB NCS Ax8 Lat4 & 106 & 81.1 & 55.3 & 63.8 \\
\hline
\end{tabular}
\caption{Electronic coupling values (meV) for hole ($\text{h}^{+}$) and excess electron ($\text{e}^{-}$) transport obtained using POD in vacuum (Vac) and condensed phase (PBC), for systems in centrosymmetric (CS) and non-centrosymmetric (NCS) alignments.}
\label{vac:con}
\end{table}

\begin{figure}[h!]
    \centering
    \includegraphics[height=0.8\linewidth]{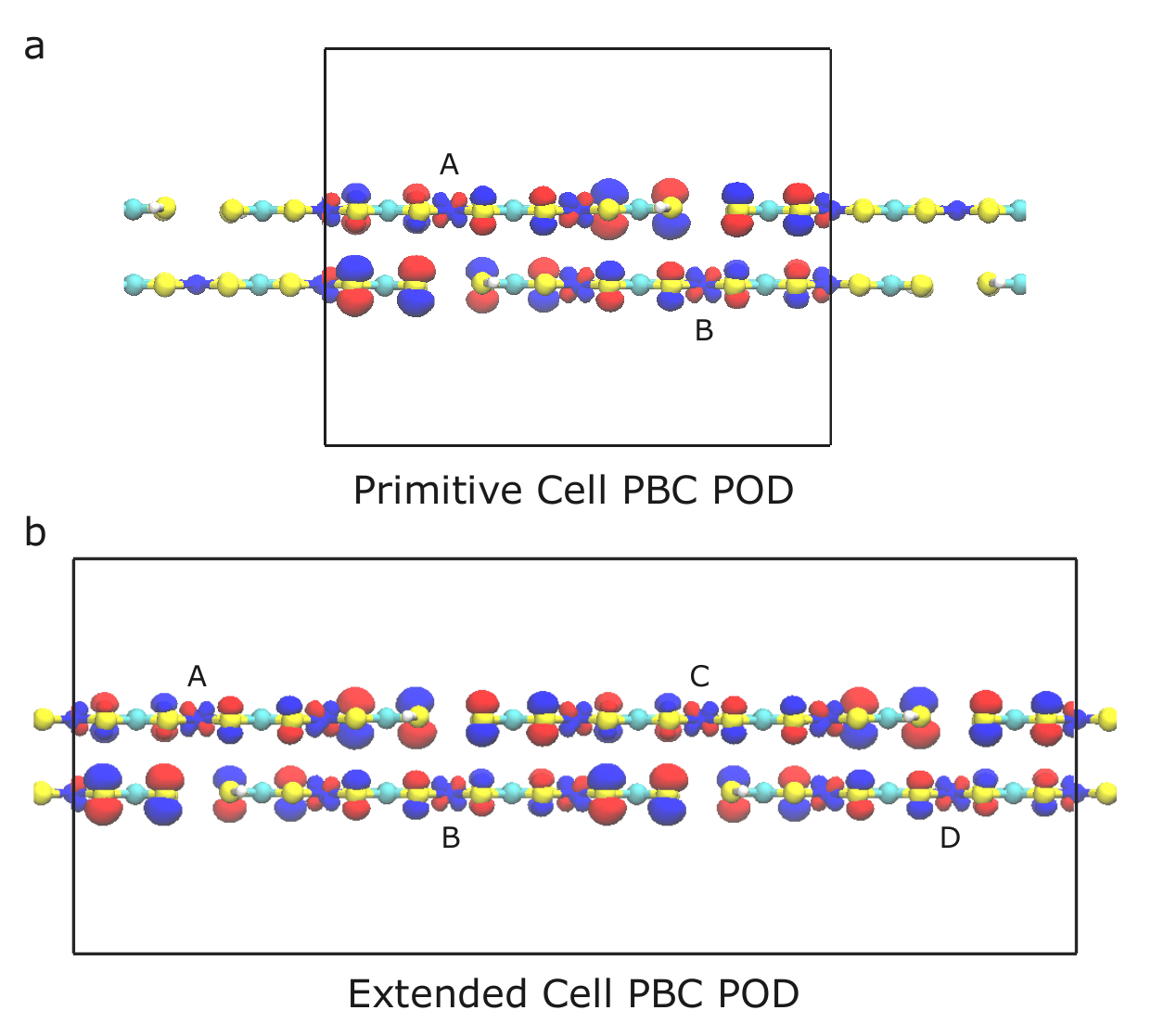} 
    \caption{\textbf{Choice of supercell for electronic coupling calculations in PBC.} In \textbf{a}, the primitive unit cell (2 molecules per cell) is shown together with the POD HOMO orbitals for molecules A and B. Neither orbital is fully represented within the unit cell; instead, the orbitals wrap across the periodic boundaries, leading to double counting of orbital overlap. An axially extended unit cell is shown in \textbf{b}, together with the POD orbitals for molecules A-D. In this case, periodic orbital wrapping no longer influences orbital overlap. Additionally, `left' (orbital B to orbital A, denoted 'L' in Table \ref{tbl:LR}) and `right' (orbital B to orbital C, denoted 'R' in Table\ref{tbl:LR}) electronic couplings can be distinguished.}

    \label{fig:primvsext}
\end{figure}

As discussed in the main text, electronic coupling is primarily dictated by orbital overlap. Hence, orbital polarization inherently influences electronic coupling values for both hole and excess electron coupling, with hole coupling being overestimated and excess electron coupling underestimated - this is reflected in Table \ref{vac:con}.
POD electronic couplings calculated in the condensed phase were done so under 1D periodic boundary conditions, periodic in the axial direction as orbital polarization is primarily along the axial direction (Figure \ref{fig:podprob}). Additionally, extension (in the axial direction) of the primitive unit cell  to include 4 monomers was necessary to avoid finite system size errors for electronic coupling calculations under periodic boundary conditions. This also distinguishes between 'left' and 'right' electronic couplings (Figure \ref{fig:primvsext}).

Interestingly we observe that for centrosymmetric configurations, 'left' and 'right' electronic coupling values are non-equivalent, with larger differences observed for larger lateral displacements. The same is not observed for non-centrosymmetric configurations, with 'left'/'right' electronic coupling differences typically being smaller than 2 meV. The 'left'/'right' non-equivalence in electronic coupling observed for CS configurations arises due to a 'left'/'right' non-equivalence in overlap environments. Evaluation of the influence of 'left'/'right' non-equivalence on charge transport is outside the current scope of investigation hence, the mean of 'left' and 'right' electronic coupling values are taken for all structures except those that distort.

\begin{table}
\begin{tabular}{ccccc}
\hline
Configuration &  L/R ($\text{h}^{+}$) & Mean ($\text{h}^{+}$)  &  L/R ($\text{e}^{-}$) & Mean ($\text{e}^{-}$)  \\
\hline
AB CS Ax8 Lat0 & 130, 168 & 149 & 172, 195 & 184\\
AB CS Ax8 Lat1 & 95.5, 154 & 125 & 159, 150 & 155\\
AB CS Ax8 Lat2 & 50.6, 112 & 81.3 & 115, 90.8 & 103\\
AB CS Ax8 Lat3 & 22.3, 105 & 63.7 & 84.9, 59.2 & 72.1\\
AB CS Ax8 Lat4 & 7.44, 114 & 60.7 & 78.5, 53.7 & 66.1\\
\hline
AB NCS Ax8 Lat0 & 98.9, 97.9 & 98.4 & 180, 181 & 181 \\
AB NCS Ax8 Lat1 & 86.8, 84.2 & 85.5 & 155, 152 & 154 \\
AB NCS Ax8 Lat2 & 66.2, 65.5 & 65.9 & 103, 100 & 102\\
AB NCS Ax8 Lat3 & 69.2, 70.6 & 69.9 & 71.5, 69.7 & 70.6 \\
AB NCS Ax8 Lat4 & 81.1, 81.1 & 81.1 & 64.1, 63.5 & 63.8 \\

\hline
\end{tabular}
\caption{Electronic coupling values (meV) for hole ($\text{h}^{+}$) and excess electron ($\text{e}^{-}$) transport obtained from calculations in PBC using POD. Both non-equivalent coupling values left (L) and right (R) and their mean are provided, see Fig. \ref{fig:primvsext} for definition of L and R.}
\label{tbl:LR}
\end{table}

\clearpage

\subsection*{Non-Centrosymmetric Plot}
\begin{figure}[h!]
    \centering
    \includegraphics[width=1\linewidth]{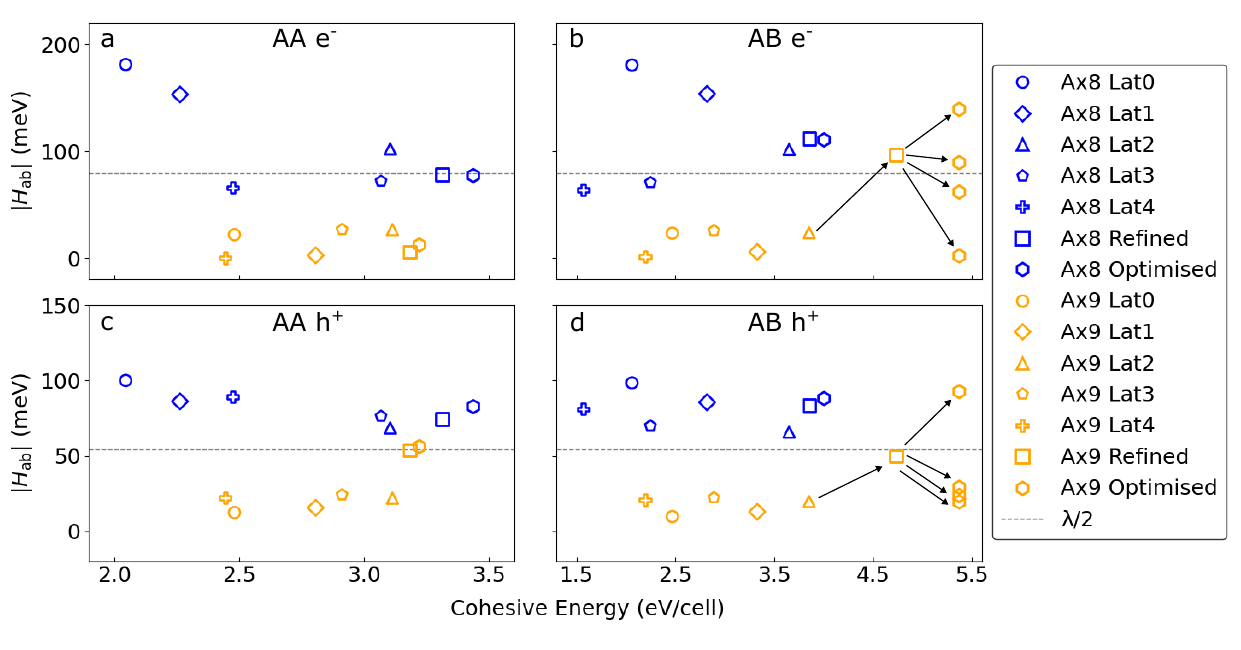} 
    \caption{\textbf{Electronic coupling versus packing stability for non-centrosymmetric nanoribbon model structures.} Electronic coupling (y axis) and cohesive energy (x axis) are plotted for centrosymmetric packing structures, with color indicating axial displacement index and marker shape indicating lateral displacement index. Refined and optimized structures are shown as squares and hexagons, respectively. AA and AB packing motifs are shown for excess electron coupling in \textbf{a} and \textbf{b}, respectively, and for hole coupling in \textbf{c} and \textbf{d}, respectively. Dashed gray lines indicate the $\lambda/2$ threshold, as in Fig.~\ref{fig:2}. Black arrows in \textbf{b} and \textbf{d} indicate the progression from 'Coarse' to 'Refined' structures (triangle to square) and from 'Refined' to 'Optimized' structures (square to hexagon) for the highest stability structures. Analogous data for centrosymmetric configurations are shown in Fig.~\ref{fig:5} in the main text.}   \label{fig:noncentro1d}
\end{figure}

\begin{table}
\begin{tabular}{ccccc}
\hline
Configuration & $\text{h}^{+}$ & $\text{e}^{-}$ \\
\hline

\hline

AA CS Ax8 Lat0 & 152 & 184 \\
AA CS Ax8 Lat1 & 125 & 154 \\
AA CS Ax8 Lat2 & 82.7 & 102 \\
AA CS Ax8 Lat3 & 70.2 & 73.0 \\
AA CS Ax8 Lat4 & 70.5 & 68.1 \\
AA CS Ax8 Refined & 77.3 & 82.7 \\
AA CS Ax8 Optimized & 76.6 & 81.5 \\
\hline
AA CS Ax9 Lat0 & 49.0 & 15.7 \\
AA CS Ax9 Lat1 & 35.0 & 19.0 \\
AA CS Ax9 Lat2 & 12.2 & 31.0 \\
AA CS Ax9 Lat3 & 12.2 & 29.3 \\
AA CS Ax9 Lat4 & 9.40 & 8.21 \\
AA CS Ax9 Refined & 43.0 & 24.6 \\
AA CS Ax9 Optimized & 43.1 & 22.0 \\
\hline
AB CS Ax8 Lat0 & 149 & 184 \\
AB CS Ax8 Lat1 & 125 & 155 \\
AB CS Ax8 Lat2 & 81.3 & 103 \\
AB CS Ax8 Lat3 & 63.7 & 72.1 \\
AB CS Ax8 Lat4 & 60.7 & 66.1 \\
AB CS Ax8 Refined & 71.0 & 80.3 \\
AB CS Ax8 Optimized & 104 & 115 \\
\hline
AB CS Ax9 Lat0 & 49.2 & 17.2 \\
AB CS Ax9 Lat1 & 38.0 & 12.6 \\
AB CS Ax9 Lat2 & 17.3 & 28.4 \\
AB CS Ax9 Lat3 & 10.1 & 28.2 \\
AB CS Ax9 Lat4 & 8.85 & 9.42 \\
AB CS Ax9 Refined & 88.4 & 107 \\
AB CS Ax9 Optimized & 0, 217, 25.1, 60.9 & 40.8, 178, 44.5, 14.8 \\
\hline

\end{tabular}
\caption{ Electronic coupling values (meV) for hole ($\text{h}^{+}$) and excess electron ($\text{e}^{-}$) transport obtained using condensed phase POD for centrosymmetric (CS) configurations plotted in Fig.~\ref{fig:4}.}
    \label{tbl:CS}

\end{table}

\begin{table}
\begin{tabular}{ccccc}
\hline
Configuration & $\text{h}^{+}$ & $\text{e}^{-}$ \\
\hline

\hline
AA NCS Ax8 Lat0 & 100 & 181 \\
AA NCS Ax8 Lat1 & 86.1 & 153 \\
AA NCS Ax8 Lat2 & 68.5 & 102 \\
AA NCS Ax8 Lat3 & 76.3 & 71.8 \\
AA NCS Ax8 Lat4 & 89.2 & 65.9 \\
AA NCS Ax8 Refined & 74.4 & 78.3 \\
AA NCS Ax8 Optimized & 82.7 & 77.1 \\
\hline
AA NCS Ax9 Lat0 & 12.4 & 21.8 \\
AA NCS Ax9 Lat1 & 15.6 & 2.46 \\
AA NCS Ax9 Lat2 & 22.0 & 26.5 \\
AA NCS Ax9 Lat3 & 24.2 & 26.6 \\
AA NCS Ax9 Lat4 & 22.2 & 0.0250 \\
AA NCS Ax9 Refined & 53.8 & 5.85 \\
AA NCS Ax9 Optimized & 56.2 & 12.2 \\
\hline
AB NCS Ax8 Lat0 & 98.4 & 181 \\
AB NCS Ax8 Lat1 & 85.5 & 154 \\
AB NCS Ax8 Lat2 & 65.9 & 102 \\
AB NCS Ax8 Lat3 & 69.9 & 70.6 \\
AB NCS Ax8 Lat4 & 81.1 & 63.8 \\
AB NCS Ax8 Refined & 83.5 & 112 \\
AB NCS Ax8 Optimized & 88.0 & 111 \\
\hline
AB NCS Ax9 Lat0 & 9.78 & 23.2 \\
AB NCS Ax9 Lat1 & 13.0 & 5.68 \\
AB NCS Ax9 Lat2 & 19.8 & 23.6 \\
AB NCS Ax9 Lat3 & 22.4 & 25.6 \\
AB NCS Ax9 Lat4 & 20.9 & 1.34 \\
AB NCS Ax9 Refined & 49.9 & 96.6 \\
AB NCS Ax9 Optimized & 29.2, 18.5, 23.8, 92.6 & 61.7, 139, 89.2, 2.00 \\
\hline
\end{tabular}
\caption{ Electronic coupling values (meV) for hole ($\text{h}^{+}$) and excess electron ($\text{e}^{-}$) transport obtained using condensed phase POD for non-centrosymmetric (NCS) configurations plotted in Fig. \ref{fig:noncentro1d}.}     
\label{tbl:NCS}

\end{table}

\end{document}